\documentclass[runningheads]{llncs}
\usepackage{graphicx}
\usepackage{color}
\usepackage{afterpage}
\usepackage{balance} 

\usepackage{graphicx}
\usepackage{cite}
\usepackage{amsmath,amssymb,amsfonts}
\usepackage{graphicx}
\usepackage{textcomp}
\usepackage{xcolor}
\usepackage{booktabs} 
\usepackage{color}
\usepackage{multirow}
\usepackage{subfigure}
\usepackage[ruled]{algorithm2e}
\usepackage[T1]{fontenc}
\usepackage{fix-cm}
\usepackage{hyperref}

\makeatletter
\newcommand{\printfnsymbol}[1]{%
  \textsuperscript{\@fnsymbol{#1}}%
}
\makeatother

\begin{document}
\title{COVID19 Tracking: An Interactive Tracking, Visualizing and Analyzing Platform}
%

\author{Zhou Yang\thanks{equal contribution}\inst{1} \and
Jiwei Xu\printfnsymbol{1} \inst{2} \and Zhenhe Pan\inst{3}
 \and Fang Jin\inst{1}}

\institute{The George Washington University\inst{1}, Long Bridge Hk Limited\inst{2}, Texas Tech University\inst{3}\\
\email{\text{zhou\_yang@gwmail.gwu.edu,ohdarling88@gmail.com, zhenpan@ttu.edu,fangjin@email.gwu.edu}}
}


%
\maketitle              
\vspace{-3mm}
\begin{abstract}
\vspace{-2mm}
The Coronavirus Disease 2019 (COVID-19) has now become a pandemic, inflicting millions of people and causing tens of thousands of deaths. To better understand the dynamics of COVID-19, we present a comprehensive COVID-19 tracking and visualization platform that pinpoints the dynamics of the COVID-19 worldwide. Four essential components are implemented: 1) presenting the visualization map of COVID-19 confirmed cases and total counts all over the world; 2) showing the worldwide trends of COVID-19 at multi-grained levels; 3) provide multi-view comparisons, including confirmed cases per million people, mortality rate and accumulative cure rate; 4) integrating a multi-grained view of the disease spreading dynamics in China and showing how the epidemic is taken under control in China. 
This demo will spur further disease spreading modeling for researchers, support decision-maker, and enrich the public awareness of the spreading situations of COVID-19 worldwide.
\end{abstract}

\vspace{-5mm}
\section{Introduction}
The outbreak of Coronavirus Disease 2019 (COVID-19) has now rapidly spread to more than 185 countries and regions, and become a global pandemic, inflicting millions of people and causing tens of thousands of deaths (as of April 10, 2020). 
Recently, a number of platforms~\cite{Chinese_web,dong2020interactive,Mistletoer_web,Tsinghua_web} about the spreading of COVID-19
have been developed in response to this increasingly exasperated global disaster.
However, most of these platforms suffer from some limitations, e.g., they either contain only the general statistics such as total reported cases in a nation, daily increment count, or only present simple visualizations of COVID-19 trend cases in a limited countries/regions.
Moreover, these existing platforms lack comparative information among different countries, which may be useful for countries that are in the different stages of the pandemic dynamics. 

In this paper, we developed an interactive platform~\footnote{https://covidtracking.app/\#tab=world-map} to track, visualize and analyze the reported cases of COVID-19 (a demo video is available\footnote{https://www.youtube.com/watch?v=NfdajZrrwtM}). It provides researchers, public health authorities and the public, with a user-friendly and comprehensive tool to track, visualize and compare the pandemic as it unfolds. 
Moreover, the platform provides a multi-view visualizations for further modeling the pandemic evolution dynamics, inferring spreading trends, and analyzing the factors that may influence the spreading dynamics of COVID-19. These insights into this challenging global emergency are extremely important to promotes the development of appropriate policies, and to contain the spreading of this pandemic. To summarize, our platform has the following contributions: 

\begin{itemize}
    \item Develop a comprehensive platform to track, visualize and compare COVID-19 cases worldwide, which would tremendously inspire disease modeling for researchers, support decision-maker and public health authorities, and enhace the public awareness of the spreading dynamics.
    \item Provide a Bird’s-eye view of the global COVID-19 dynamics and allow users to view the COVID-19 trends for each country and state/province at multi-grained levels. 
    \item Present a multi-dimension comparison among countries that are severely affected by the pandemic.
    \item Show the full evolution path of the pandemic in China, which could be beneficial for counties that haven't contained the disease spreading yet.
\end{itemize}

\vspace{-5mm}
\section{Platform Framework}
\begin{figure*}[thbp]
    \centering
    \includegraphics[width=1.2\linewidth]{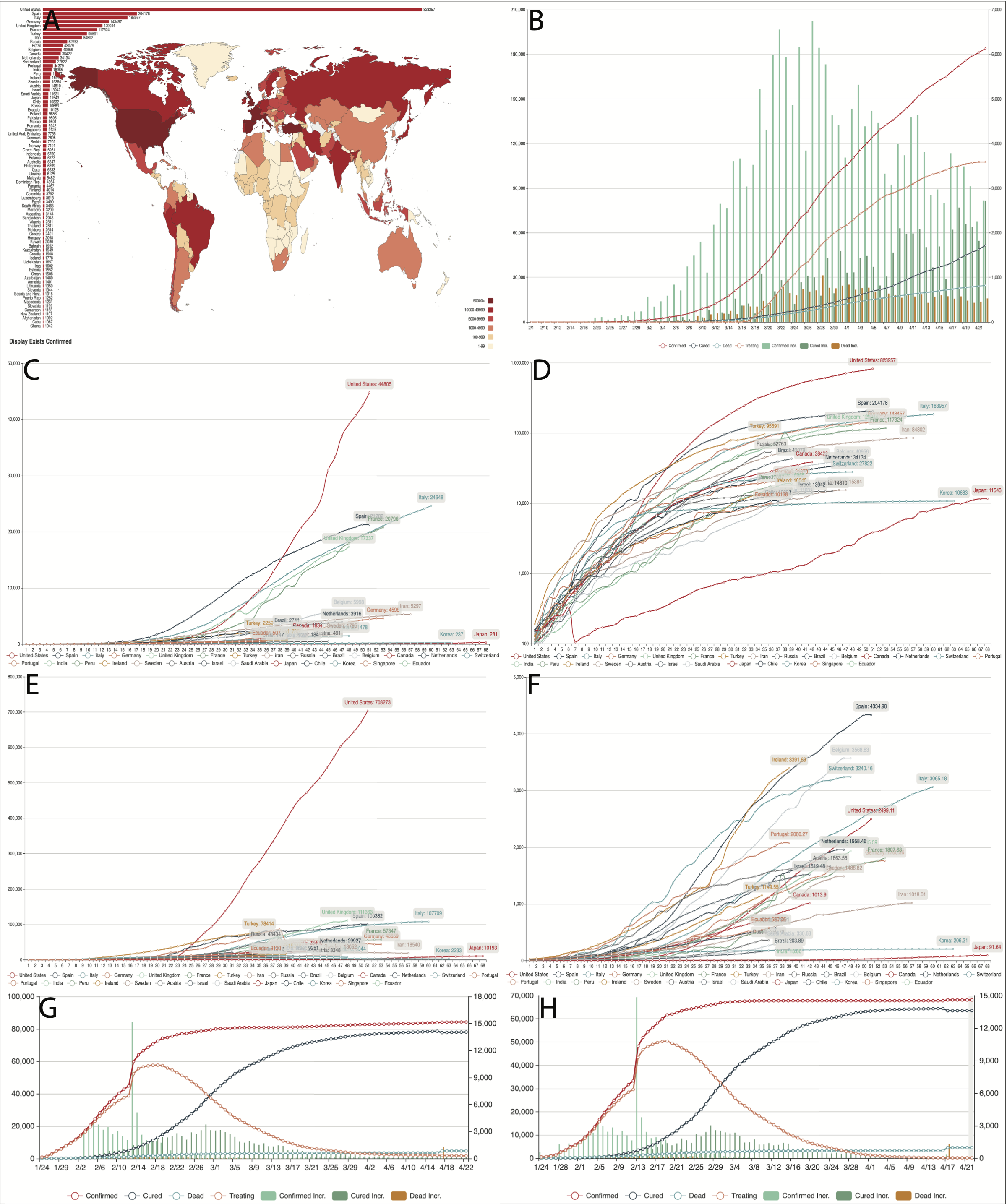}
    \caption{Landing page of the COVID-19 platform.}

    \label{fig:world_map}
\end{figure*}

In this section, we present the platform we built and pinpoint each component inside. The landing page of this platform is shown in Figure~\ref{fig:world_map}.


\vspace{-2mm}
\subsection{Dataset and Data Flow}
The primary data source is DXY, a GitHub repository~\footnote{https://github.com/BlankerL/DXY-COVID-19-Data, https://ncov.dxy.cn/ncovh5/view/pneumonia, https://covidtracking.com/} that is maintained by many volunteers, which aggregates multiple data sources, such as local media, government reports, and health organizations, to provide cumulative COVID-19 cases in near real-time. 
\vspace{-2mm}
\subsection{Global Heatmap}
As shown in part \textbf{A} of Figure~\ref{fig:world_map}, the COVID-19 heatmap shows the number of countries affected, the total number of cases, death toll, and etc. The darker the color is, the larger the accumulated case number this country/region has. If the mouse hangs over a county/region, it will highlight the contour of this country/region. 
\vspace{-2mm}
\subsection{National Trends}
We group nations into subgroups by their geographical locations and visualize information, such as total reported cases, cured cases, death counts, the total number of people under treatment, daily new confirmed cases, daily cured cases, and daily deaths. 
We show the visualization of the national trend of Italy in part \textbf{B}  
as an example. The curves marked by circles use y-axis on the left, and the bar charts share the y-axis on the right. Moreover, we also allow users to input the nation or city name to search the related visualization information.



\vspace{-2mm}
\subsection{Multi-view Comparisons}
We provide multi-view comparisons among countries, including comparisons on mortality rate in part \textbf{C}, total confirmed cases in part \textbf{D},
and existing confirmed cases in part \textbf{E}, confirmed cases per million people in part \textbf{F}.
All those visualizations are designed to allow users to select countries for comparisons. The visualization of total confirmed cases provides users the perspective to retrospect back, and users can choose a specific country to zoom in. 



\vspace{-2mm}
\subsection{Multi-grained Spreading Dynamics in China}
We detailed the accumulated reported cases, daily increment, daily cured cases, and daily deaths in China as part \textbf{G} shows.
We also present the disease spreading at province and city levels. The motivation is that different provinces or cities may have distinct evolution patterns, and hence a multi-grained view would unveil more valuable information. 
Part \textbf{H} uses data collected in Hubei province only since a big portion of the confirmed cases have Hubei travel history in China. By showing this multi-grained information, we believe the users can uncover more information underneath.




\vspace{-3mm}
\section{Conclusions}
We develop an interactive platform to track, visualize and analyze reported cases of COVID-19 worldwide, providing platform users with multi-view and multi-grained insights into the pandemic dynamics. Platform users can select countries and information levels they are interested in. At a selected nation level, the platform provides users with the latest COVID-19 cases, deaths, recoveries, daily confirmed case increment, daily deaths increment, and daily cured case increment.
This platform will tremendously help researchers with the disease spreading modeling, support decision-maker, and enrich the public awareness of the spreading of COVID-19 worldwide. 
\vspace{-3mm}
\bibliographystyle{splncs04}
\bibliography{reference}

\end{document}